# Evidence of self-organization in time series of capital markets

Leopoldo Sánchez-Cantú[1*], Carlos Soto-Campos[**], Andriy Kryvko[*]

*ESIME, Instituto Politécnico Nacional, Ciudad de México, México
**Universidad Autónoma del Estado de Hidalgo, Pachuca, México

**HIGHLIGHTS**
- We study the dynamics of prices in multiple equity market indices.
- The decrease in stock prices (draw-downs) were identified as units of study.
- A range of price differentials larger than a critical level conforms to a power law.
- This level is interpreted as a phase transition into a self-organized system.

**ABSTRACT**

A methodology is developed to identify, as units of study, each decrease in the value of a stock from a given maximum price level. A critical level in the amount of price declines is found to separate a segment operating under a random walk from a segment operating under a power law. This level is interpreted as a point of phase transition into a self-organized system. Evidence of self-organization was found in all the stock market indices studied but in none of the control synthetic random series. Findings partially explain the fractal structure characteristic of financial time series and suggest that price fluctuations adopt two different operating regimes. We propose to identify downward movements larger than the critical level apparently subject to the power law, as self-organized states, and price decreases smaller than the critical level, as a random walk with the Markov property.

## 1. Introduction

The analytical-reductionist methodology that has traditionally addressed the description and study of stock market price fluctuations [9], [152], [69], [108], [109], [115], [39], [40], [120], [121], [102], [62] has left out of its scope of observation, therefore without explanation, some of the most interesting features of the phenomena since they don't fit a normal Gaussian distribution and do not meet the principles of a Wiener-type random walk with the Markov property generated by supposedly rational agents as the unbiased aggregated response to the random flow of external information [96].

Instead of considering these characteristics as anomalies [31], [83], [122], we propose that heavy-tailed distributions [84], [85], the presence of clusters of high volatility alternating with periods of low volatility [123], [124], [38] [17], [97], the non-stationarity of statistical parameters [137], [93], [29], [68], the characteristic multifractal structure of stock market time series [112], [89], [20], [27], [75], [101], the recurring crises that stock markets have suffered since the seventeenth century [70], [134], and a fundamental instability inherent to financial markets [98] are structural processes that need to be addressed and explained instead of being swept under the rug.

Several authors have tested the hypothesis that price fluctuations have short, medium and long-term memory [88], [54], [76], [77], [77], [72]. However, findings have been mixed, especially for terms longer than a few hours or days. On the flip side, long-term memory has consistently been demonstrated in the volatility of time series [36], [17], [37].

---

[1] Corresponding author.
E-mail address: polo.antares@gmail.com (L. Sánchez-Cantú)

This paper addresses price fluctuations with a systemic outlook to explain the alleged "anomalies" as emergent phenomena [3], [5], [6]. To this aim, recurrent downward movements of asset prices have been selected as observable units of study in a large sample of time series (30 cases) of equity market indices.

## 2. Methodology

A methodology was developed to identify in a series of daily closing prices each downward movement to a bottom from a recent peak, followed by a rebound back to the previous ceiling, or back to the highest value registered in the previous six months, whichever was reached first. These cumulative negative returns or downfalls then became units of study. Using the series of negative returns as observables, we explored the possibility of identifying a range within the space of states in which a variable that corresponded to the lowest value accumulated during each downfall could be explained as a process that follows a power law.

Data sets of downfalls of 30 international equity market indices (7 regional international, 5 American, 4 Latin American, 4 from emerging European countries, 5 from developed European countries and 5 from Asian countries) were ordered by size. The absolute value of each downfall (ordinates) was plotted against the place it occupied by size (abscissas) in a log-log scale.

Kurtosis of the progressive sets of absolute downfalls of each index were calculated, anchoring each series at the smallest decrease in value. Increasingly larger absolute decreased values were incorporated one by one until the largest was reached. The level of the one decrease from which the set of downfalls smaller to it had a kurtosis closest to zero was identified as the cutoff point or critical level. The set of events with kurtosis closest to zero is thus compatible with a normal density distribution.

The cutoff point was recognized as the critical level of phase transition separating two regimes of operation. The lower (mesokurtotic) segment of smaller price decreases, allegedly operating under a random regime, while the upper (leptokurtotic) segment of downfalls larger than the critical level can be explained as following a power law. This latter set of larger losses is incompatible with a normal density distribution. We propose to consider this set of larger declines in value purportedly generated by a process under a self-organized regime.

## 3. Computation of parameters

We obtained a time series of daily closing prices/values, $c_i$, of each stock index from the earliest date available to us, $c_0$, down to the most recent date at the time the study was made, $c_n$ (initial and final dates of each series are shown on Table 2). We then completed the following operations to estimate all the parameters used in this study:

- Log returns $r_i$, of price series were estimated with $r_i = \ln(c_i/c_{i-1})$. Standard deviations, $S_r$, were calculated as usual. Kurtosis, $K_r$, of daily returns were calculated as excess-kurtosis with…

$$K = \frac{1}{N}\sum_{i=1}^{N} \frac{(x_i - \bar{x})^4}{\sigma^4} - 3 \qquad (1)$$

…so the normal (mesokurtotic or Gaussian) value is 0. Larger values are considered leptokurtic or leptokurtotic [151]. The total number of log returns for each index are labeled $N_r$.

- The set of daily values of the maximum closing price of the previous six months, $c_{Max}$, was generated together with the closing price, $c_i$, of every series (Figure 1, A and B).

- We calculated a series of daily differences between $c_i - c_{Max}$ to measure each decrease in value of the series $c_i$ below the series $c_{Max}$. These differences also show the rebounding of $c_i$ back to the level of $c_{Max}$. Series of values, $d_i = c_i - c_{Max}$, were generated (C in Figure 1). Each one of these movements is named a draw-down. A complete draw-down was considered as the set of negative values, $d_i < 0$, located between two alternating points where $d_i = 0$.

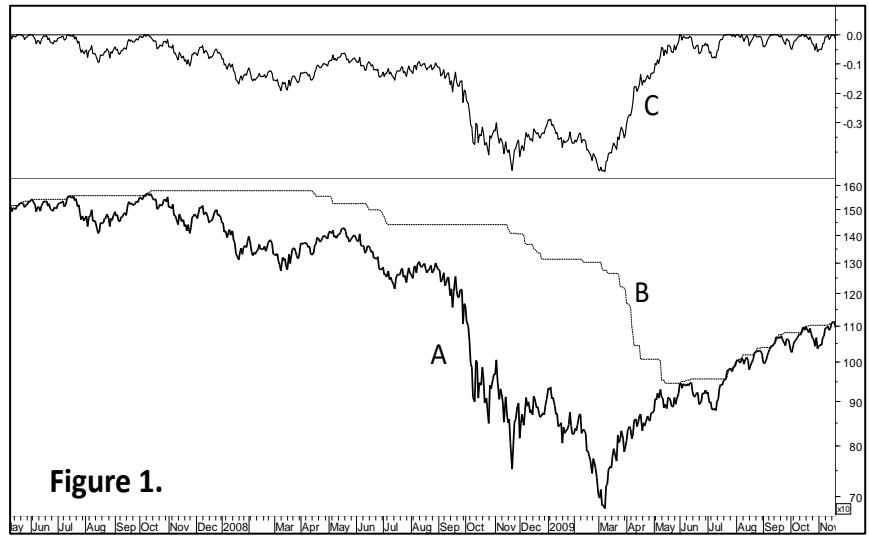

**Fig. 1.** Nominal value of the S&P500 Index, $c_i$, (A), the maximum closing level of the last 6 months, $c_{Max}$, (B) (inferior panel, semi-log scale). Difference, $c_i - c_{Max}$, (C) (superior panel, arithmetic scale).

- The most negative value of each $d_i$ series (the bottom of every fall) was recorded (such $d_i$ value was named $d_{max}$). For further analysis, the absolute value of each $d_{max}$ is identified as an observable unit of study $x_i$ (Figure 2). The complete set of $x_i$ values for each index was $N_x$.

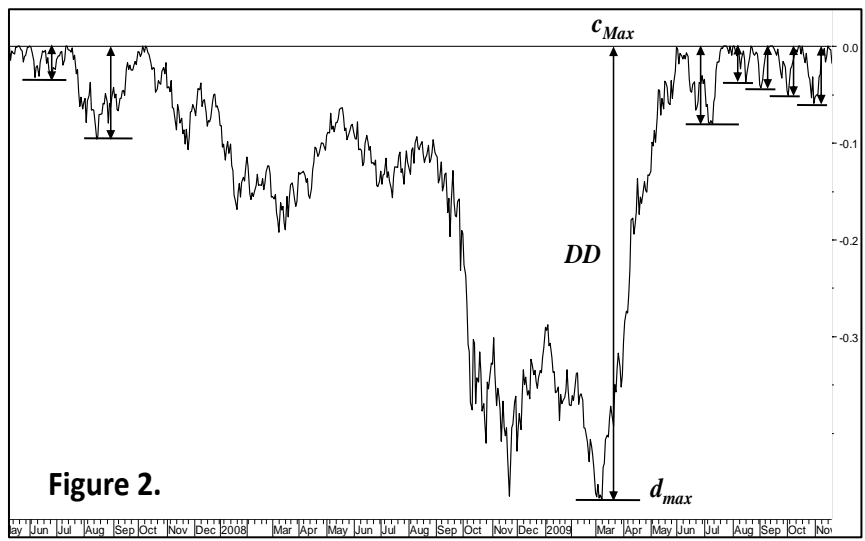

**Fig. 2.** Difference between the daily nominal value of the S&P500 index compared to the maximum closing value of the previous 6 months $c_i - c_{Max}$. The arrows show how to identify the maximum value of a price drop $d_{max}$, followed by a complete recovery of the loss until it reaches back the reference value $c_{Max}$. DD stands for draw-down.

- Maximum draw-down values, $x_i$, were ordered from the largest, corresponding to the largest decline, $x_{Max}$, to the smallest, corresponding to the shallowest decline, $x_0$.
- This generated a scatterplot of $x_i$ values (ordinates) against the cumulative place they occupied (abscissa) in a log-log scale (Figure 3).

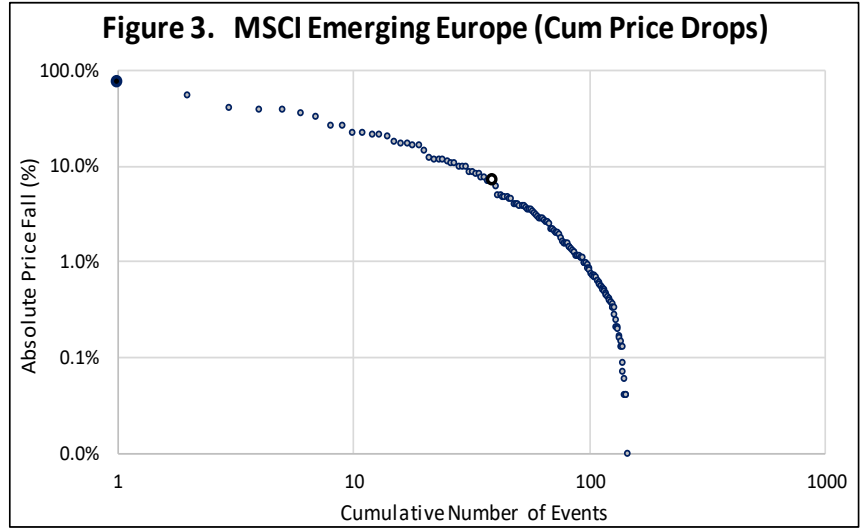

**Fig. 3**. Scatterplot of individual price drops ($d_{max}$ in absolute percentage value). Ordered from the deepest (largest) to the shallowest (smallest), in a log-log scale (critical data-point, $x_{min}$, in white).

- Kurtosis, $K_x$, of the complete series of declines, $x_i$, for each index was calculated. Also calculated were the kurtosis of the sets of cumulative values, $K_s$, anchored at $x_{Max}$ down to $x_0$. The same occurred with the sets of cumulative values, $K_i$, anchored at $x_0$ up to $x_{Max}$ (see Table 1).
- There followed an identification of the inferior segment of declines in value with a kurtosis value, $K_i$, closest to zero. This segment is thus considered mesokurtotic. The $x_i$ value at the cutoff point of this segment is labeled $x_{min}$ from here on.
- Kurtosis, $K_s$, of the superior segment (from $x_{Max}$ down to $x_{min}$) was also calculated. This segment was found to be leptokurtotic for all indices. The point of minimum fall, $x_{min}$, in this set, equivalent to the phase transition point, is the point at which the regime changes from being random (declines of less than $x_{min}$ dimension) to a self-organizing regime (declines greater than $x_{min}$ dimension), $x_{min}$ being the critical level which separates them.
- Declines with values ranging from $x_{Max}$ to $x_{min}$ were selected and depicted on a log-log scale (log of the cumulative number of events against log of the size of the fall). A power regression line for this set of events was drawn and its coefficient of determination $R^2$ was recorded (see Figure 4). The number of events in this set was labeled $N_s$.
- The value of the exponent, $\alpha$, of the regression and the value of the standard error of the exponent, $\sigma$, were calculated using the following formulas [106], [30]:

$$\alpha = 1 + N_s \left[ \sum_{i=1}^{n} \ln \frac{x_i}{x_{min}} \right]^{-1} \tag{2}$$

$$\sigma = \frac{\alpha - 1}{\sqrt{N_s}} \tag{3}$$

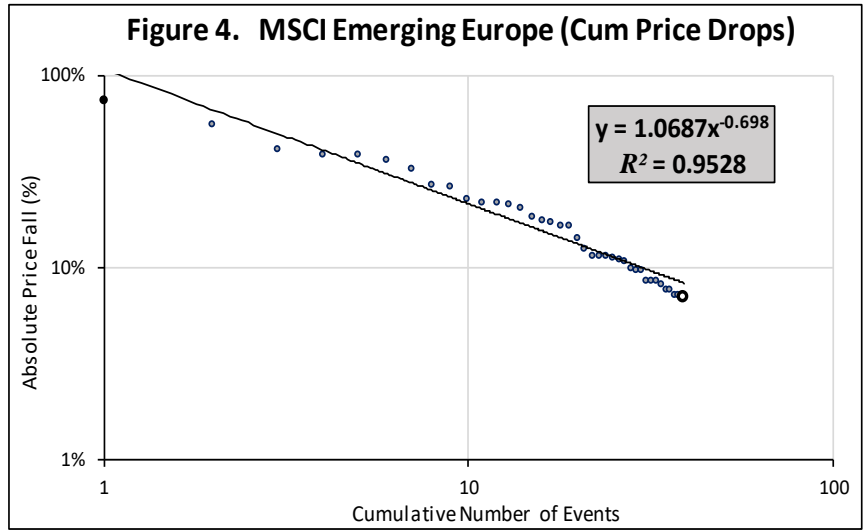

**Fig. 4.** Segment of the complete set of $x_i$ falls (gray circles) in a log-log scale. Includes data from the deepest fall ($x_{Max}$, in black), down to that fall at the critical point ($x_{min}$, in white). The power regression line, its formula and the $R^2$ Coefficient (inside the box), are included.

The next 12 parameters were calculated for each one of the 30 stock market indices studied:
- Total number of price decreases registered in the series, $N_x$.
- Number of events (cases) in the upper segment, $N_s$.
- Number of events (cases) in the lower segment, $N_i$.
- Percentage that $N_s$ represents of $N_x$.
- Kurtosis, $K_x$, of the complete set of downward tendencies.
- Kurtosis, $K_s$, of the upper (superior) downward tendencies.
- Kurtosis, $K_i$, of the lower (inferior) downward tendencies.
- Value of $x_{Max}$.
- Value of $x_{min}$.
- $R^2$ value of power regression line of upper set of downward tendencies.
- Value of $\alpha$ exponent of power regression line.
- Standard error, $\sigma$, of the $\alpha$ exponent of the power regression line.

## 4. Results

Table 1 shows the most relevant results. The general characteristics of the financial time series (number of data for each series, standard deviation and kurtosis of daily log-returns, initial and final dates) appear in Table 2. The 30 stock market indices analyzed are briefly described in the Appendix.

The method we used to identify each observation unit (each dip) yielded events whose amount depended on the length of history of the index in days of operation. The correlation coefficient between both quantities (days of operation vs. degree of loss registered) is 0.992 ($R^2$ = 0.945). The average number of losses registered in all 30 indexes is 323, with a minimum of 111 for Colombia (2001-2015) and a maximum of 1,147 for the DJIA (1897-2015). The average proportion of events or downturns larger than the critical point is 23.68 % (ranging from 15.95% to 32.20%).

The average excess-kurtosis of the complete sets of price declines, $K_x$, (formula (1)) is 22.17 (7.925 to 64.709). The average excess-kurtosis of the set of these declines above the critical point, $K_s$, is 6.029 (1.76 to 21.552) whereas the average excess-kurtosis of the inferior set of declines, $K_i$, is 0.001 (-0.295 to 0.151).

The average value of the deepest loss, $x_{Max}$, for all series is -55.91% (-38.91% to -77.62%). The average value of the loss at the critical point, $x_{min}$, is -4.46% (-1.73% to -8.70%) and the average coefficient of determination, $R^2$, of the power regression is 0.9584 (0.9110 to 0.9834).

| Table 1. | a | b | c | d | e | f | g | h | i | j | k | l |
|---|---|---|---|---|---|---|---|---|---|---|---|---|
| | $N_{tot}$ | $N_{sup}$ | $N_{inf}$ | % Sup | $K_{tot}$ | $K_{sup}$ | $K_{inf}$ | $x_{Max}$ | $x_{min}$ | $R^2$ | $\alpha$ | $\sigma$ |
| MSCI ACWI | 295 | 57 | 238 | 19.32% | 39.276 | 12.174 | 0.053 | -50.99% | -3.27% | 0.9744 | 2.200 | 0.162 |
| MSCI WI | 326 | 79 | 247 | 24.23% | 40.576 | 15.989 | 0.045 | -49.54% | -3.36% | 0.9666 | 2.171 | 0.132 |
| EM EUROPE | 145 | 39 | 106 | 26.90% | 12.890 | 4.408 | 0.049 | -73.80% | -6.92% | 0.9528 | 2.207 | 0.193 |
| EM ASIA | 259 | 59 | 200 | 22.78% | 23.506 | 6.084 | -0.005 | -61.00% | -2.48% | 0.9686 | 2.864 | 0.192 |
| EM LATAM | 267 | 71 | 196 | 26.59% | 16.447 | 3.898 | 0.021 | -68.08% | -4.32% | 0.9381 | 2.087 | 0.129 |
| EuroStox 50 | 139 | 34 | 105 | 24.46% | 18.379 | 3.363 | -0.013 | -46.24% | -4.16% | 0.9698 | 2.320 | 0.226 |
| StoxxEuro 600 | 358 | 72 | 286 | 20.11% | 28.189 | 4.984 | -0.016 | -43.66% | -2.55% | 0.9635 | 2.079 | 0.127 |
| S&P500 | 696 | 170 | 526 | 24.43% | 31.442 | 12.022 | -0.002 | -46.91% | -2.81% | 0.9808 | 2.393 | 0.107 |
| WILSHIRE 5K | 240 | 69 | 171 | 28.75% | 64.709 | 21.552 | 0.151 | -56.64% | -1.73% | 0.9712 | 2.103 | 0.133 |
| DJIA | 1147 | 266 | 881 | 23.19% | 22.962 | 6.327 | -0.003 | -53.57% | -3.12% | 0.9471 | 2.147 | 0.070 |
| NASDAQ C. | 452 | 105 | 347 | 23.23% | 28.815 | 9.108 | 0.086 | -54.99% | -3.37% | 0.9659 | 2.172 | 0.114 |
| RUSSELL 2K | 263 | 54 | 209 | 20.53% | 21.868 | 4.508 | 0.008 | -54.93% | -4.28% | 0.9659 | 2.226 | 0.167 |
| FRANCE | 247 | 58 | 189 | 23.48% | 17.583 | 3.656 | 0.037 | -43.38% | -4.61% | 0.9683 | 2.372 | 0.180 |
| GERMANY | 525 | 125 | 400 | 23.81% | 22.209 | 5.636 | -0.003 | -51.70% | -3.61% | 0.9583 | 2.249 | 0.112 |
| ITALY | 172 | 51 | 121 | 29.65% | 11.915 | 2.208 | -0.004 | -56.09% | -3.31% | 0.9242 | 1.959 | 0.134 |
| U. KINGDOM | 257 | 41 | 216 | 15.95% | 19.506 | 3.666 | -0.036 | -39.59% | -5.48% | 0.9766 | 2.607 | 0.251 |
| SWITZERLAND | 266 | 54 | 212 | 20.30% | 20.637 | 2.939 | -0.295 | -41.01% | -3.62% | 0.9567 | 2.269 | 0.173 |
| JAPAN | 436 | 91 | 345 | 20.87% | 21.945 | 4.107 | -0.002 | -50.94% | -3.38% | 0.9834 | 2.211 | 0.127 |
| HONG KONG | 408 | 100 | 308 | 24.51% | 18.655 | 5.125 | -0.001 | -72.14% | -5.50% | 0.9607 | 2.258 | 0.126 |
| AUSTRALIA | 359 | 83 | 276 | 23.12% | 31.418 | 10.688 | 0.005 | -50.09% | -3.01% | 0.9582 | 2.187 | 0.130 |
| TAIWAN | 432 | 93 | 339 | 21.53% | 19.634 | 4.747 | -0.037 | -73.64% | -5.26% | 0.9436 | 2.140 | 0.118 |
| INDIA | 282 | 73 | 209 | 25.89% | 11.177 | 3.094 | 0.008 | -52.02% | -6.18% | 0.9607 | 2.439 | 0.168 |
| RUSSIA | 138 | 37 | 101 | 26.81% | 12.884 | 4.901 | 0.005 | -76.83% | -8.09% | 0.9692 | 2.158 | 0.190 |
| TURKEY | 236 | 76 | 160 | 32.20% | 7.925 | 1.869 | 0.054 | -59.13% | -5.70% | 0.9110 | 2.048 | 0.120 |
| HUNGARY | 172 | 41 | 131 | 23.84% | 12.561 | 2.719 | 0.020 | -58.16% | -5.69% | 0.9464 | 2.108 | 0.173 |
| POLAND | 199 | 50 | 149 | 25.13% | 16.833 | 5.537 | -0.008 | -67.57% | -5.62% | 0.9551 | 2.254 | 0.177 |
| MEXICO | 439 | 98 | 341 | 22.32% | 28.136 | 7.953 | -0.042 | -77.62% | -4.78% | 0.9749 | 2.240 | 0.125 |
| BRAZIL | 226 | 38 | 188 | 16.81% | 18.380 | 2.815 | -0.003 | -61.42% | -8.70% | 0.9435 | 2.470 | 0.238 |
| CHILE | 202 | 64 | 138 | 31.68% | 12.083 | 3.023 | 0.006 | -38.91% | -2.12% | 0.9376 | 1.963 | 0.120 |
| COLOMBIA | 111 | 20 | 91 | 18.02% | 12.530 | 1.760 | -0.042 | -46.78% | -6.75% | 0.9584 | 2.257 | 0.281 |
| AVERAGE | 323 | 76 | 248 | 23.68% | 22.169 | 6.029 | 0.001 | -55.91% | -4.46% | 0.9584 | 2.239 | 0.157 |
| MAX | 1147 | 266 | 881 | 32.20% | 64.709 | 21.552 | 0.151 | -38.91% | -1.73% | 0.9834 | 2.864 | 0.281 |
| MIN | 111 | 20 | 91 | 15.95% | 7.925 | 1.760 | -0.295 | -77.62% | -8.70% | 0.9110 | 1.959 | 0.070 |
| | $N_{tot}$ | $N_{sup}$ | $N_{inf}$ | % Sup | $K_{tot}$ | $K_{sup}$ | $K_{inf}$ | $x_{Max}$ | $x_{min}$ | $R^2$ | $\alpha$ | $\sigma$ |

**Table 1.** Values of the main measurements carried out to the time series of each market index are as follow: a) total number of drops registered [$N_{tot}$]; b) number of drops larger than the critical point $x_{min}$ [$N_{sup}$]; c) number of drops smaller than the critical point $x_{min}$ [$N_{inf}$]; d) percentage of drops larger than the critical point $x_{min}$ as compared to the total number of drops [% Sup]; e) excess kurtosis of the complete set of drops [$K_{tot}$]; f) excess kurtosis of the set of drops larger than the critical point [$K_{sup}$]; g) excess kurtosis of the set of drops smaller than the critical point [$K_{inf}$]; h) largest drop $x_{Max}$ registered in each series [$x_{Max}$]; i) critical point $x_{min}$ in each series of drops [$x_{min}$]; j) coefficient of determination $R^2$ of the power regression line for the set of drops larger than the critical point [$R^2$]; k) exponent $\alpha$ of the power regression line [$\alpha$]; l) standard error $\sigma$ of the exponent of the power regression line [$\sigma$].

It was considered incorrect to estimate the scaling exponent using the power regression line adjustment because, as Goldstein et al [53] highlighted, this procedure is systematically biased and induces errors; hence, it is not reliable. The method Newman [106] and Clauset [30] proposed was used instead. The formulas are annotated at number 11 of Computation of

Parameters as in formulas (2) and (3). The average value of the $\alpha$ exponent is 2.239 (1.959 to 2.864), and the standard error of that exponent is, on average, 0.157 (0.070 to 0.281).

Excess kurtosis of the set of downward tendencies smaller than the critical level (not included in Figure 4) is approximately zero (actual values of all series are depicted on Table 1, col. g), while the density distribution of the set of values larger than the critical level included in the graph corresponds to the leptokurtic segment (Table 1, col. f).

Figure 5 shows the critical level, $x_{min}$, of each index by groups: regional indices, USA indices, Emerging Markets Europe, Developed Countries Europe, Latin America Markets and Asia Markets. Note that Latin America indices are the most disperse while the USA and Developed Europe indices have the least disperse values. Furthermore, the average value of the critical point is more negative for the Latin America indices (-5.58%) and Emerging Europe (-6.27%), is intermediate for the Asia indices (-4.466%) and Developed Europe (-4.12%) and is less negative for the Regional indices (-3.86%) and the USA indices (-3.062%). This may be related to the degree of efficiency (randomness) of the series, and should be addressed.

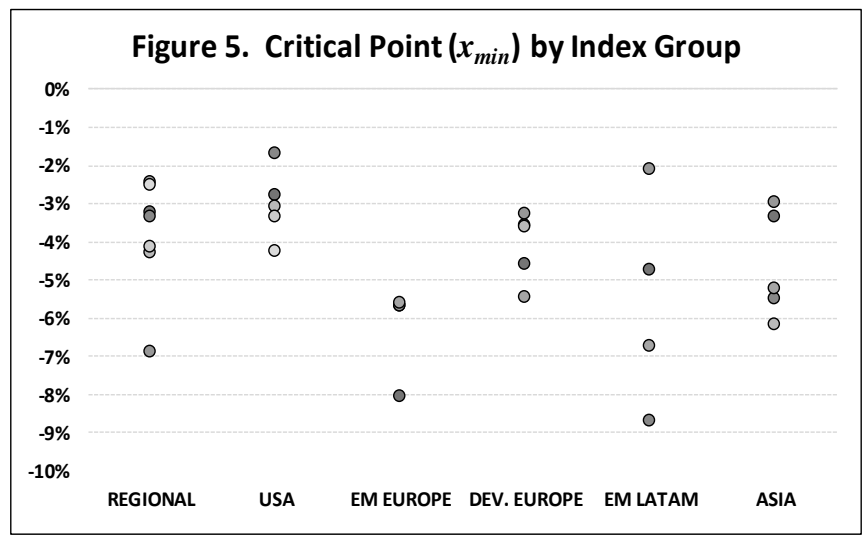

**Fig. 5**. Level of those price drops, $x_{min}$, that identify the critical point of the grouped indexes in Regional [REGIONAL = 7] USA indexes [USA = 5], Emerging Europe [EM EUROPE= 4], Developed Europe [DEV. EUROPE = 5], Latin America indexes [EM LATAM = 4], Asian Markets indexes [ASIA = 5].

## 5. Discussion

The task of explaining and modeling price fluctuations in financial markets, one of the most complex phenomena we can imagine, has been challenging. It was bravely undertaken by giants with the stature of Bachelier, Working, Savage, Marschak, Samuelson, Sharpe and others [9], [152], [45], [96], [108], [115], [120]. With a reductionist view, based on strongly restrictive and unrealistic assumptions, their work, however, fell on fertile soil whose fruits could be harvested in abundance throughout the second half of the 20th century, creating a theoretical framework that is elegant, pure, beautiful, timeless and balanced. Nevertheless, in addition to other authors [138], [117], [73], [149], [77], [126], [25], [63], [127], [128], [51], [113], [65], [66], [67], [146], [147], [129], [130], [142], [143], [144], [145], [125], [52], [78], [48], [1], [55], [56], [64], [119], we believe the time has come to liberate most of those assumptions and to get closer to the phenomenon as it is: organic, dynamic, unstable, rough, discontinuous, partially self-generated and diffuse. Perhaps the state of affairs today is too complex for that old model.

| Table 2. | a | b | c | d | e |
|---|---|---|---|---|---|
|  | $N_r$ | $S_r$ | $K_r$ | First Date | Last Date |
| MSCI ACWI | 7,058 | 0.909% | 8.374 | 1/3/1988 | 3/13/2015 |
| MSCI WI | 7,748 | 0.923% | 11.334 | 6/10/1985 | 3/13/2015 |
| EM EUROPE | 4,974 | 1.886% | 10.597 | 1/2/1995 | 3/13/2015 |
| EM ASIA | 7,088 | 1.291% | 6.583 | 12/31/1987 | 3/13/2015 |
| EM LATAM | 4,126 | 1.717% | 9.268 | 1/1/1988 | 2/26/2015 |
| EuroStox 50 | 4,126 | 1.504% | 4.348 | 1/22/1999 | 3/13/2015 |
| StoxxEuro 600 | 7,268 | 1.115% | 7.156 | 12/31/1986 | 3/13/2015 |
| S&P500 | 16,418 | 0.970% | 27.675 | 1/3/1950 | 4/6/2015 |
| WILSHIRE 5K | 6,526 | 1.121% | 8.805 | 3/31/1989 | 2/26/2015 |
| DJIA | 32,180 | 1.097% | 31.420 | 02/01/1897 | 3/4/2015 |
| NASDAQ C. | 11,139 | 1.240% | 9.888 | 2/5/1971 | 4/7/2015 |
| RUSSELL 2K | 6,949 | 1.317% | 8.640 | 9/10/1987 | 4/7/2015 |
| FRANCE | 4,126 | 1.401% | 5.288 | 7/9/1987 | 3/17/2015 |
| GERMANY | 13,941 | 1.229% | 7.620 | 10/1/1959 | 3/16/2015 |
| ITALY | 6,410 | 1.456% | 5.248 | 12/29/1989 | 4/7/2015 |
| U. KINGDOM | 7,821 | 1.102% | 8.447 | 4/2/1984 | 3/17/2015 |
| SWITZERLAND | 6,725 | 1.166% | 7.777 | 7/1/1988 | 4/7/2015 |
| JAPAN | 10,541 | 1.304% | 10.072 | 1/5/1970 | 3/18/2015 |
| HONG KONG | 11,181 | 1.875% | 30.253 | 12/1/1969 | 4/2/2015 |
| AUSTRALIA | 8,916 | 0.990% | 85.615 | 12/31/1979 | 4/7/2015 |
| TAIWAN | 13,251 | 1.549% | 3.015 | 1/4/1967 | 4/7/2015 |
| INDIA | 8,213 | 1.651% | 5.891 | 4/3/1979 | 4/7/2015 |
| RUSSIA | 4,377 | 2.754% | 14.496 | 9/22/1997 | 4/7/2015 |
| TURKEY | 6,574 | 2.717% | 4.096 | 1/4/1988 | 4/7/2015 |
| HUNGARY | 6,053 | 1.663% | 10.941 | 1/2/1991 | 4/7/2015 |
| POLAND | 5,524 | 1.922% | 7.438 | 4/16/1991 | 4/7/2015 |
| MEXICO | 10,027 | 1.727% | 20.468 | 1/3/1975 | 3/5/2015 |
| BRASIL | 5,432 | 2.367% | 10.182 | 4/14/1993 | 4/2/2015 |
| CHILE | 6,304 | 0.817% | 6.818 | 1/2/1990 | 4/2/2015 |
| COLOMBIA | 3,362 | 1.329% | 12.488 | 7/3/2001 | 4/9/2015 |
| AVERAGE | 8,479 | 1.470% | 13.341 |  |  |
| MAX | 32,180 | 2.754% | 85.615 |  |  |
| MIN | 3,362 | 0.817% | 3.015 |  |  |
|  | $N_r$ | $S_r$ | $K_r$ | First Date | Last Date |

**Table 2.** $N_r$ = Total number of daily log-returns of each original series. $S_r$ = Standard Deviation of total daily log-returns of each original series. $K_r$ = Kurtosis of total daily log-returns of each original series. *First Date*, date of the first data in the series, corresponds to $c_0$ of the series $c_i$. *Last Date*, date of the last data in the series, corresponds to $c_n$ of the series $c_i$.

The fact that the financial world is in a permanent state of innovation and conceivably complexification, it is also inherent in its nature to undergo recurrent crises of endogenous origin that should be faced once and for all. Concepts such as holism, entropy, dissipative systems, unbalanced states, self-organization, positive feedback loops and non-linearity, characteristics of the complexity approach, have been gradually introduced into the theoretical discourse of economics and finance [3], [7], [5], [6], [58], [60], [44], [22], [15], [107], [135], [154]. These concepts offer useful alternatives which may allow us to liberate restrictions that are not only unbelievable but sometimes rather absurd. Perfect rationality of economic agents, a selfish interest in maximizing personal utility, and the supposed homogeneity of expectations are just a few examples of assumptions that might hinder the progress in this field.

Doubtlessly the behavioral patterns at the level of each agent that makes decisions is important; however, it is insufficient to explain the aggregated result or the supposedly coordinated activity of all participants, each one with his or her own peculiarities, while sharing a place and a role in the same system at the same time. Modeling the stock market phenomena, we seek to build a bridge from the micro-level, i.e. from the clever but fallible single agent, with relatively simple and rather stereotypical strategies, responsible for the input that goes into the system (trading in the marketplace), to the output observed at the macro level, that is, price fluctuations with their stylized characteristics.

We consider such an output to be the result not only of the cumulative effect of the parts, likewise as something richer and more organic, i.e., something that confers novel properties on the system. From a systemic perspective, that "something" we propose may be called emergent properties.

New paths recently explored to model the stock markets are related to two main branches: A) The identification and description of the stylized micro-structure of financial time series and the fractal properties of such series and B) Agent-Based Simulations create synthetic markets that follow the seminal segregation model of Thomas Schelling [118]. The pioneering work of Benôit Madelbrot excels in the field of fractality, together with a mounting group of authors devoted to it [89], [112], [92], [93], [149], [136], [91], [8], [27], [2], [68], [27], [12], [71], [18], [23], [99], [100], [134], [103], [104], [114], [101]. Through the creation of stock market simulations, individual agents are given heterogeneous characteristics and allowed to freely interact in a synthetic environment. These interactions reliably generate price fluctuations portraying the stylized characteristics found in real life markets [59], [7], [110], [80], [42], [43], [50], [35], [74], [141], [81], [61], [43].

## 6. The power law and self-organization

In an attempt to relate the heavy-tailed distribution with the fractal properties of price fluctuations, the clusters of high volatility, the capacity of self-organization and the emergent properties of complex adaptive systems, we explored the possibility of finding a distribution form known as Power Law or Zipf's Law in the deepest fluctuations or tails of falling prices given the ubiquity of this kind of distribution in phenomena with statistical properties such as those found in financial time series [150].

In 1949, the American philologist George Zipf discovered that when the American corporations listed on the stock market were ordered by size (market capitalization value), the size $S_n$ of the $n^a$ largest company was inversely proportional to the place it occupied in the series in an approximate way to the form $S_n \sim 1/n^a$.

Previously, Zipf had found that the frequency distribution of words in a text followed the same rule [155]. Nowadays this distribution form is known as the Zipf's Law. Half a century before Zipf's publication (in 1949), the Italian engineer, sociologist, philosopher and economist Vilfredo Pareto, described a pattern known as *tail function*, applicable to several social and physical phenomena whose cumulative distribution function of continuous variables can't be distinguished from Zipf's proposal for discrete ones [111]. Another of the few differences between Zipf's and Pareto's proposals is that Zipf did his graphs with the $x$ value (the measured variable) on the horizontal axis and the probability of appearance $P(x)$ on the vertical axis, while Pareto did it the other way around. The latter is the way we do it in this paper.

Zip's Law is a special case of the Pareto distribution. Both forms of distribution have interesting statistical properties to be contrasted with our results. Firstly, the heavy tails distribution reflected the leptokurtic distribution, notable in all time-series studied here (Table 1, column f). It is said that these processes are dependent on the Power Law because the probability of obtaining a specific value in any investigated parameter varies inversely as the exponent of such value.

The phenomena that have these properties are not properly represented by a typical value or arithmetic average because the extreme cases can deviate from the media by several orders of magnitude [106], [30]. Examples of natural and social phenomena with this property are the distribution of wealth among members of a community [111], [21], [26], the Gutenberg-Richter Law of frequency of occurrence and intensity of earthquakes [57], the intensity of solar flares [79], the size and frequency of moon craters [105], the size of cities as to number of inhabitants and to their frequency [47], the formation of random networks of several kinds [14], and many others [94].

It is alerted that very few forms of distribution of real world phenomena follow the Power Law in the complete range of values that they may adopt, particularly for small values of the measured variable. In fact, for any positive value of exponent $\alpha$, the function $p(x) = Cx^\alpha$ diverges as $x$ tends to zero. Therefore, the distribution must deviate from the form of the Power Law under a certain minimum value $x_{min}$. The form of the Power Law appears as a straight line only for values greater than $x_{min}$ [106].

From a geometrical perspective, a phenomenon organized under the Power Law has fractal properties, that is, within a certain range of values there is self-similarity at different scales; hence it is said that they have scale invariance. It is characteristic of price fluctuations that the extreme events lack a scale or typical value around which the individual cases are concentrated in their magnitude and duration. Generally, this is a consequence of the central limit theorem for processes free of scale in which a Levy random walk replaces the Brownian movement [19].

In this paper, we have found an inverse relation between the size of price decreases and their frequency of presentation through their exponents in such a way that, if a graph is built of such relationships on an arithmetic scale, the distribution adopts a curve in the shape of "J", which approaches both orthogonal axes asymptotically. When the graph is made on a log-log scale, the distribution forms a straight line with a negative slope. In both cases, all segments of the curve are similar if they are represented in the proper scale.

How come an earthquake, the size of moon craters, the human dwellers in a city or the accumulation of wealth in a family are related to price reductions in the stock market? And, why should they have similar statistical and geometrical properties?

As Sornette and Cauwles suggest, maybe the kind of phenomena that behave in this way can have catastrophic results, apparently detonated by trivial exogenous events that occasionally reach a tipping point since during the phases of apparent equilibrium and tranquility, the necessary conditions for the so-called *avalanche,* accumulate. For such a feared outcome to happen, a critical value in state variables must be reached before the phenomena self-organize and change their regime [135]. The proverbial straw that breaks the camel back is an example in which, after a paused and peaceful process and almost without warning, some relevant variable reaches a critical level and manifests as a crisis in the form known as a crossover or bifurcation.

In the case of decreased prices in the stock market, while the system is still under the random regime, small changes generated by the response of some investors to exogenous information, perceived as negative, are easily absorbed by the contingent of optimistic participants who may be analyzing the same information but in a longer (or shorter) time frame. Perhaps they conclude that the recent descent in prices creates a favorable condition to increase their positions. However, if the downhill movement of prices continues, the sequence of successive perturbations increases the tension generated in the system following new small downward impulses. At a certain point, the pressure upon the contingent of buyers eventually overrides their capacity to absorb the increasing bidding until a further decrease in prices generates a change of regime (bifurcation) in which rather than buyers, new sellers who wish to get rid of their positions to stop their growing losses are attracted. Thus, a positive feedback loop is built: lowered prices attract more sellers whose bidding presses prices downwards in a vicious circle that generates a selling crisis characteristic of a market sell-off.

Furthermore, as prices continue to fall, the process could be accelerated as the mandatory selling of positions reaches a threshold in the form of margin-calls, or due to risk management criteria, which trigger stop loss signals. These two possibilities are clear examples of positive feedback mechanisms capable of accentuating the lowering of prices with total disregard to exogenous information or to disparities of the market price with the supposed intrinsic value of assets anticipated by the efficient market hypothesis. Instead, a self-organized system that activates the reinforcement of price trends emerges. It is precisely this self-organized phenomenon in a regime, which reinforces itself that we believe is happening during the phase of dropping prices that are larger than the critical point identified in our model as $x_{min}$.

There is an endogenous process resulting from an internal re-structuring which depends on the new relationship among agents conforming the system and their answers to information derived from the system itself. This expression of the phenomena is an intimate arrangement in the balance of perturbing and repairing forces, or, using systemic language, the homeostatic mechanisms are surpassed in such a way that the system adopts a new regime that instead of being a stabilizer of its output it becomes an amplifier.

From a mathematical point of view, this observation is not a surprise. It can be understood as the generic behavior of a dynamic system. In accordance with the general theorems of the bifurcation theory, there is only a finite number of ways in which a system can change its regime; it is a change that happens suddenly, not progressively [135].

In his excellent review of the subject, Mark Newman from The University of Michigan, Ann Arbor, discusses the statistical distribution of the Power Law and describes several mechanisms proposed to explain its occurrence. These include: 1) the mechanism known as highly optimized tolerance (HOT) of Carlson and Doyle, 2) the Sneppen and Newman mechanism focused on the behavior of agents under stress, 3) the process of Udly Yule in critical phenomena and 4), the concept of a self-organized critical state anticipated by Per Bak [106].

The Highly-Optimized Tolerance of Carlson and Doyle [28] proposes that in natural and in human systems organized to offer a robust performance, regardless of the uncertainties in the environment, an exchange is generated among the revenue, the cost of the resources and the tolerance to risk, which leads to highly optimized designs that are predisposed to sporadic events of great magnitude. The main characteristics of the systems in a HOT state include: 1) high efficiency performance and robustness to uncertainties to which they are designed, 2) hypersensitivity to design flaws or unforeseen perturbations, 3) structurally, not generic, specialized configurations, and 4) being subject to the Power Law. Classic examples are forest fires and other phenomena based on the percolation model [28].

Another mechanism, mathematically equivalent to that of Carlson and Doyle's, is *coherent noise,* proposed by Sneppen and Newman. In this mechanism, a certain number of agents or species are subject to stress of different kinds for which each agent has a threshold. Above that level of stress, the agent will be eliminated, or the species will be extinguished. Extinct species are replaced by new species with stress thresholds randomly selected. The net result is a system that self-organizes to a final state where many of the species will have high levels of stress tolerance. This type of phenomenon shows reorganization events whose size is distributed in accordance to the Power Law through many decades. Additionally, the system shows after-shock events with the same distribution. The authors propose that under the action of a slow local force, some systems with short-range interactions may organize to a critical state without needing to fine tune some parameters [131].

The Yule process is a mechanism known as "the rich get richer" or Gibrat's Law, a principle of cumulative advantage or preferential selection. Here, an alternative that occupies a prominent place as a possible choice will have a higher probability of being chosen; thus, it will have an amplifying effect mathematically demonstrated as a Power Law distribution in its tail [106], [153]. This mechanism is probably adequate to explain that the size of companies and their frequency in a market, the frequency of words in a text or the number of assets a family possesses, show a heavy tailed distribution and follow the Power Law. However, to explain the processes that determine that a price drop of stock assets which surpasses a certain critical level extends the descent in a self-organized form with characteristics such as those shown in this paper, we prefer the method described by Per Bak, that is, a self-organization to a critical state or Self Organized Criticality [11].

According to this model, when the falling prices reach a critical level, a progressive recruitment of sellers who try to limit their losses by getting rid of their positions starts. Each agent will have his/her own pre-existing conditions (or "schema", as Murray Gell-Mann tags them [49]), such as exposure to the market, previous accumulated returns, degree of exposure and risk control policies and tolerance or aversion to risk, that will determine the threshold level necessary to reach before triggering a selling process. Under Self Organized Criticality, the reaction of sellers is triggered independently of the exogenous information delivered at that moment and of the previous expectations agents have. Instead, it will be determined solely by the endogenous information derived from a price fall of certain magnitude. Once this critical point is reached, the system diverges from its previous trajectory and begins a different dynamic process, as Balcilar et al. have suggested in relation to the herd behavior [13].

Bouchaud has suggested that this erratic dynamic of the markets is mainly of endogenous origin. He attributes this to a market that operates in a regime of manifest evanescent liquidity, but, at the same time, high latent liquidity, which explains its hypersensitivity to fluctuations and identifies a dangerous positive feedback loop arranged by the spread in prices of supply (bid) and demand (ask) and volatility that may lead to a crisis of micro-liquidity and huge jumps in prices [24].

## 7. Practical applications

We propose four ways in which these concepts may be directly applied to the practice of finance. First, Zipf's distribution in the size of companies has an important consequence in building alleged efficient portfolios [95] given the impossibility of making an adequate diversification of specific risk when there is dominance of some small number of companies of great size (capitalization) in a market. Therefore, its effect in the supposed "market portfolio" can't be diversified even if the total number of companies included in the sample is large. This is considered a particular risk factor named Zipf's factor [82].

Second, it is possible to explain the phases of high volatility as an emergent process that results from a regime in which agents' expectations diverge about their returns on investments, or about the target prices of assets. In other words, their beliefs are dispersed in an exaggerated way compared with the baseline heterogeneity typically observed under normal circumstances. At the same time, the dispersion in the timeframe adopted to observe, analyze and react to exogenous and endogenous events, becomes more homogeneous. We suggest that as the timeframes in which agents make decisions become shorter and more homogeneous, and as their expectations become more heterogeneous and disperse, the market becomes less stable and more vulnerable to both endogenous and exogenous perturbations. We still have to design the appropriate tests to identify these characteristics in empirical series or replicate these mechanisms with agent-based simulation models to prove our suggestion.

Third, as proposed by Sornette and Scheffer, as we better understand the way in which markets self-organize in successive layers, and how the downfalls in prices are generated, we may recognize the mechanisms that determine the avalanches, develop indicators that allow us to evaluate their presence, find early signs that the system is reaching a critical level of phase transition and, perhaps, mitigate the local or general effects of the disruptive process [132], [133], [116].

And last, having found that the exponent of the regression line of the tails is smaller than 3 (media of 2.239 and range from 1.959 to 2.864), we can predict the probability and magnitude of large negative fluctuations observed in capital markets once the threshold or critical point is exceeded. We are currently developing this methodology.

It could be that from the rich interaction among components of the system, and the non-linear processes they generate, an inherently unpredictable dynamism results. However, what could interest us is predicting the big bifurcations or breaking points towards extreme events happening when critical levels are surpassed in a set of parameters from which an explosion to the infinite is conveyed upon a usually stable variable. We need to investigate how patterns of great scale and catastrophic nature evolve; we suppose there are growing levels of auto-correlation of the relevant variables, formed from processes occurring on a smaller scale. The critical points identified in the present paper suggest the possibility that, on a larger scale, other events that better explain the great stock market breakdowns may be identified.

## 8. Conclusions

In this paper, we design a new model that may explain price fluctuations as a process with heavy tails distribution that follows a Power Law. We assume a non-stationary phenomenon with periods of high volatility alternating with others of low volatility. To this end, we have considered the statistical and geometric properties of financial time series as emergent phenomena resulting from the joint activity and interaction of autonomous agents, heterogeneous in many aspects, operating in an unstable, high complexity context and resulting in the appearance of trends in price fluctuations and distortions towards extreme states.

We have developed a method to identify the critical point of phase transition in which the random regime proposed by the classical paradigm gives way to a self-organized regime. We have found that the downward movements of prices may be explained as alternating states between periods compatible with a random walk with i.i.d. properties, and periods subject to a self-organized emergent regime that may explain the presence of heavy tails.

Specifically, we have identified that price drops that go deeper than a critical point may be represented as a phenomenon under the Power Law. From this characteristic identified we have made a tentative explanation of how it is generated and what potential consequences such a proposed mechanism may produce to generate a phenomenon with the demonstrated statistical and geometrical characteristics observed. This may explain, at least partially, the fractal structure with self-affinity and scale invariance observed in financial time series.

Our findings suggest the presence of medium-term memory, tentatively due to the effect of positive feedback loops [4] which also identify a probable generation mechanism of clusters with high volatility. We propose to identify descendent movements subject to the Power Law as self-organized states of the kind described by Per Bak as Self-Organized Criticality [10], [11], [46].

Upon the bases established in this paper, we need to define a mechanism for estimating the probability of occurrence of decreases in stock prices larger than the critical point, through the definition of the proper characteristics of each series. This would result in the first practical application of these new concepts.

**Appendix**

The 30 indexes investigated are as follows:

- MSCI ACWI (Morgan Stanley Capital International All Countries World Index) includes shares of 2,446 firms from 46 countries and captures returns in 23 Developed Economies and 23 Emerging Markets countries. It represents all economic sectors worldwide.
- MSCI WI (Morgan Stanley Capital International World Index), includes shares of 1,613 firms from 23 Developed Economy countries. It represents all economic sectors.
- MSCI EM EUROPE (MSCI Emerging Markets Europe Index) captures representation across 6 Emerging Markets countries from Europe (Russia, Turkey, Poland, Greece, Hungary, Czech Republic). It has 84 constituents.
- MSCI EM ASIA (MSCI Emerging Markets Asia Index) represents 534 firms from 8 Emerging Market countries from Asia (China, India, Indonesia, Korea, Malaysia, the Philippines, Taiwan and Thailand).
- MSCI EM LATAM (MSCI Emerging Markets Latin America Index) with 118 constituents from 5 Emerging Markets countries from Latin America (Brazil, Mexico, Chile, Colombia and Peru).
- Stoxx Europe 600 represents 600 components of large, mid and small capitalization companies across 18 European countries (Austria, Belgium, Czech Republic, Denmark, Finland, France, Germany, Greece, Ireland, Italy, Luxembourg, the Netherlands, Norway, Portugal, Spain, Sweden, Switzerland and the United Kingdom).
- Euro Stoxx 50, the leading blue-chip index from the Eurozone, covers 50 stocks from 12 countries (Austria, Belgium, Finland, France, Germany, Greece, Ireland, Italy, Luxembourg, the Netherlands, Portugal and Spain).
- CAC 40 Index, France (40 stocks)
- FTSE 100 Index, Great Britain (100 stocks)
- DAX Index, Germany (30 stocks)
- FTSE MIB Index, Italy (40 stocks)
- Swiss Market Index, Switzerland (20 stocks)
- MICEX Index, Russia (50 stocks)
- BISE National 100 Index, Turkey (100 stocks)
- BSE Index, Hungary (13 stocks)
- WIG Index, Poland (342 stocks)
- S&P500 (500 stocks, Large-Caps, USA)
- DJIA (30 stocks, Large-Caps, USA)
- NASDAQ Composite (2,976 stocks, USA)
- Russell 2000 (2000 stocks, Small-Caps, USA)
- Wilshire 5000 Index (3,698 stocks, USA)
- IPC, Mexico (35 stocks)
- BOVESPA, Brazil (71 stocks)
- IGPA, Chile (102 stocks)
- IGBC, Colombia (34 stocks)
- NIKKEI 225, Japan (225 stocks)
- HANG SENG, Hong Kong (50 stocks)
- All Ordinaries Index, Australia (498 stocks)
- TAIEX Index, Taiwan (786 stocks)
- S&P BSE SENSEX Index, India (30 stocks)